\documentclass[superscriptaddress,notitlepage,preprintnumbers,amsmath,10pt,aps,pra]{revtex4-1}
\usepackage{graphicx}
\usepackage{color}
\usepackage{subfigure}
\def\change#1{\textcolor{black}{#1}}
\def\red#1{\textcolor{black}{#1}}

\begin{document}
\title{Experimental high-dimensional two-photon entanglement and \\violations of generalised Bell inequalities}

\affiliation{SUPA, School of Engineering and Physical Sciences, Heriot-Watt University, Edinburgh EH14 4AS, United Kingdom}
\affiliation{Department of Physics and Astronomy, SUPA, University of Glasgow, Glasgow G12 8QQ, United Kingdom}

\author{Adetunmise C. Dada}
 \email{acd8@hw.ac.uk}
\affiliation{SUPA, School of Engineering and Physical Sciences, Heriot-Watt University, Edinburgh EH14 4AS, United Kingdom}
 
\author{Jonathan Leach}
\affiliation{Department of Physics and Astronomy, SUPA, University of Glasgow, Glasgow G12 8QQ, United Kingdom}

\author{Gerald S. Buller}
\affiliation{SUPA, School of Engineering and Physical Sciences, Heriot-Watt University, Edinburgh EH14 4AS, United Kingdom}

\author{Miles J. Padgett}
\affiliation{Department of Physics and Astronomy, SUPA, University of Glasgow, Glasgow G12 8QQ, United Kingdom}

\author{Erika Andersson} 
\affiliation{SUPA, School of Engineering and Physical Sciences, Heriot-Watt University, Edinburgh EH14 4AS, United Kingdom}

\maketitle
\twocolumngrid
\textbf{Quantum entanglement~\cite{PhysRev.47.777,schro1935} plays a vital role in many quantum information and communication tasks~\cite{Nielsen2000}. Entangled states of higher dimensional systems are of great interest due to the extended possibilities they provide. For example, they allow the realisation of new types of quantum information schemes that can offer higher information-density coding and greater resilience to errors than can be achieved with entangled two-dimensional systems (see~\cite{PhysRevA.69.032313} and references therein). Closing the detection loophole in Bell test experiments is also more experimentally feasible when higher dimensional entangled systems are used~\cite{PhysRevLett.104.060401}. We have measured previously untested correlations between two photons to experimentally demonstrate high-dimensional entangled states. We obtain violations of Bell-type inequalities generalised to $d$-dimensional systems~\cite{PhysRevLett.88.040404} with up to 
$d=12$. Furthermore, the violations are strong enough to indicate genuine 11-dimensional entanglement.
Our experiments use photons entangled in orbital angular momentum (OAM)~\cite{PhysRevA.45.8185}, generated through spontaneous parametric down-conversion (SPDC)~\cite{PhysRevLett.75.4337,PhysRevA.69.023811}, and manipulated using computer controlled holograms.}

Quantum information tasks requiring high dimen\-sional bipartite en\-tan\-gle\-ment in\-clu\-de tele\-por\-ta\-tion us\-ing qu\-dits~\cite{PhysRevLett.70.1895,doi:827398}, generalised dense coding (i.e., with pairs of entangled $d$-level
systems)~\cite{PhysRevLett.69.2881}, and some quantum key distribution protocols~\cite{PhysRevLett.67.661}. More generally, schemes like quantum secret sharing~\cite{PhysRevA.59.1829}, and measurement based quantum computation~\cite{PhysRevA.68.022312}, apply multi-particle entanglement. 
These are promising applications, especially in view of recent progress in the development of quantum repeaters (see~\cite{citeulike:3168245} and references therein).
However, practical applications of such protocols are only conceivable when it is possible to experimentally prepare, and moreover, detect high-dimensional entangled states. Therefore, the ability to verify high-dimensional entanglement between physical qudits is of crucial importance.  Indeed, much progress has generally been made on the generation and detection of high-dimensional entangled states~(please see~\cite{citeulike:6916025} and references within).

In this Letter, we report the experimental investigation of high-dimensional, two-photon entangled states. We focus on photon OAM entangled states generated by SPDC, and demonstrate genuine high-dimensional entanglement using violations of generalised Bell-type inequalities~\cite{PhysRevLett.88.040404}. Previously, qutrit Bell-type tests have been performed using photon OAM to verify 3-dimensional entanglement (see ~\cite{PhysRevLett.89.240401} and references within).
In addition to testing whether correlations in nature can be explained by local realist theories~\cite{PhysRevLett.49.91}, the violation of Bell-type inequalities may be used to demonstrate the presence of entanglement. 
Bell-type experiments have been performed using two-dimensional subspaces of the OAM state space of photons~\cite{Jleach2009,PhysRevA.81.043844} and experimentalists have demonstrated 2-dimensional entanglement using up to twenty different 2-dimensional subspaces~\cite{1367-2630-11-10-103024}.  Careful studies have also been carried out to describe how specific detector characteristics bound the dimensionality of the measured OAM states in photons generated by SPDC using Shannon dimensionality~\cite{PhysRevLett.101.120502}. 

Our experimental study of high\--dimensional entanglement is based on the theoretical work of Collins {\it et al.}~\cite{PhysRevLett.88.040404}, which was applied in experiments for qutrits encoded in the OAM states of photons~\cite{PhysRevLett.89.240401,1367-2630-8-5-075}. 
We encode qudits using the OAM states of photons, with eigenstates defined by the azimuthal index $\ell$. These states arise from the solution of the paraxial wave equation in its cylindrical co-ordinate representation, and are the Laguerre-Gaussian modes $LG_{p,\ell}$, so called because they are light beams having a Laguerre-Gaussian amplitude distribution. 

\begin{figure*}[t]
\centerline{\includegraphics[width=1\textwidth]{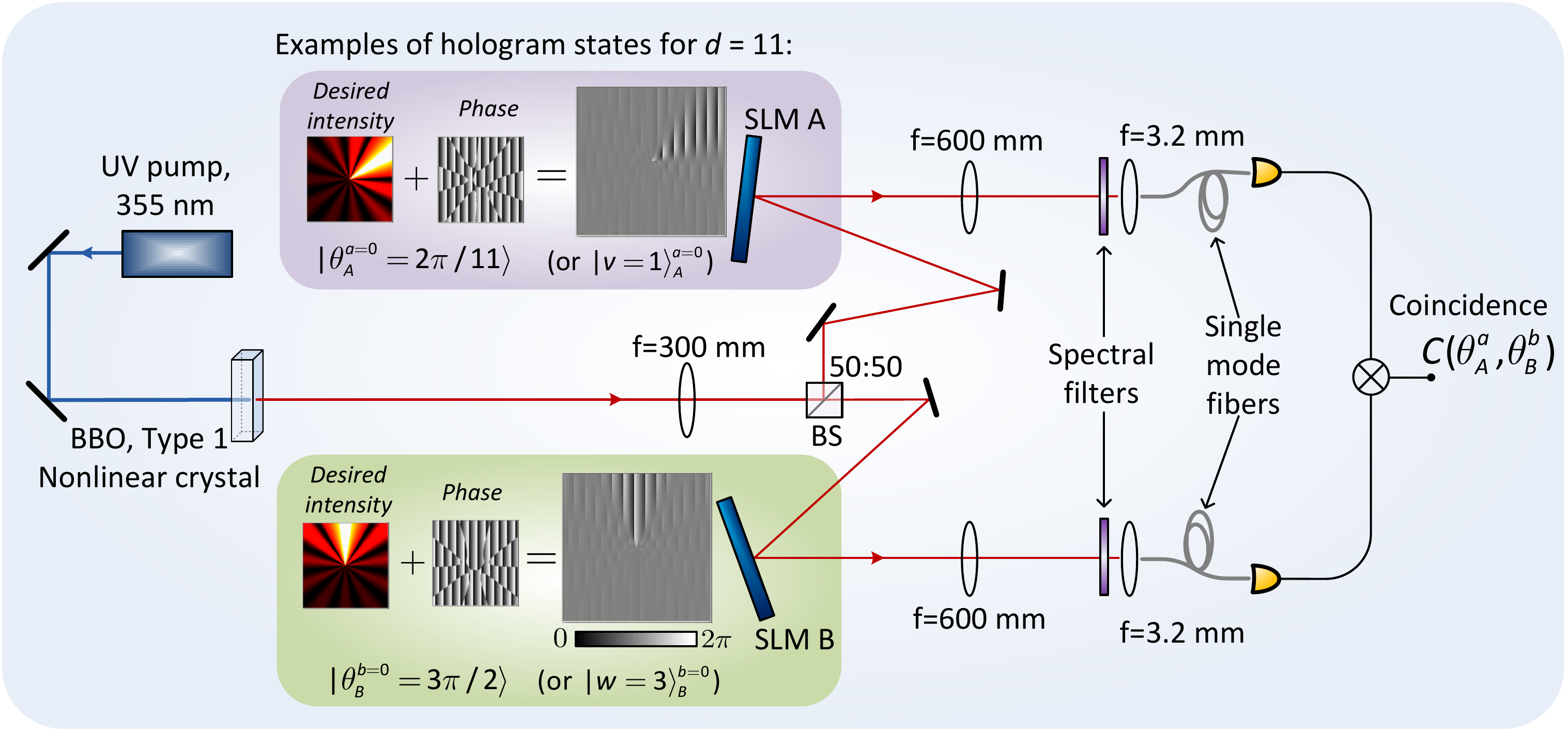}}
\caption{Schematic of experimental setup for violations of Bell-type inequalities. $C(A_a=v,B_b=w)$ or $C(\theta_A^a,\theta_B^b)$ is the coincidence count rate when SLM A is in state $|v\rangle_a^A$ or $|\theta_A^a\rangle $  and SLM B is in state $|w\rangle_b^B$ or $|\theta_B^b\rangle $ respectively.}
\label{fig:setupoam}
\end{figure*}

In our setup (Fig.~\ref{fig:setupoam}), OAM entangled photons are generated through a frequency degenerate type-I SPDC process, and the OAM state is manipulated with computer controlled spatial light modulators (SLM) acting as reconfigurable holograms. 
Conservation of angular momentum  ensures that if the signal photon is in the mode specified by ${\left| \ell \right\rangle }$, the corresponding idler photon can only be in the mode ${\left| -\ell \right\rangle }$. 
Assuming that angular momentum is conserved~\cite{PhysRevA.69.023811}, a pure state of the two photon field produced will have the form
 \begin{equation}
\left| \Psi  \right\rangle  = \sum\limits_{\ell =  - \infty }^{\ell = \infty } {c_\ell \left| \ell \right\rangle _A \otimes \left| { - \ell} \right\rangle _B }
 \label{eq:spdcstate}, 
 \end{equation}
where subscripts $A$ and $B$ label the signal and idler photons respectively, $\left| {c_\ell} \right|^2$ is the probability to create a photon pair with OAM $\pm\ell\hbar$ and ${\left| \ell \right\rangle }$ is the OAM eigenmode with mode number $\ell$. 

Collins {\it et al.}~\cite{PhysRevLett.88.040404} showed that, for correlations which can be described by theories based on local realism~\cite{PhysRev.47.777}, a family of Bell-type parameters $S_d$ satisfy the inequalities
\begin{equation}
\label{eq:s}
S_d^{(\text{local realism})} \leq 2, ~{\rm for~all}~d\geq 2.
\end{equation}
Alternatively, if quantum mechanics is assumed to hold, then the violation of an inequality of type~(\ref{eq:s}) indicates the presence of entanglement. $S_d$ can be expressed as the expectation value of a quantum mechanical observable, which we denote as $\hat{S}_d$. The expressions for $S_d$, $\hat{S}_d$ and the operators $\hat{S}_2$, $\hat{S}_3$ are provided in Sec. SI of the Supplementary Information. 

The parameters $S_d$ are calculated using coincidence probabilities for measurements made locally by two observers, Alice and Bob, on their respective subsystems, which in our case are the signal and idler photons from the SPDC source. Alice's detector has two settings labelled by $a\in\{0,1\}$ with $d$ outcomes for each setting, and similarly for Bob's detector with settings $b\in\{0,1\}$. The measurement bases corresponding to the detector settings 
of Alice and Bob are defined as
\begin{align}
|v\rangle_a^A=&\tfrac{1}{\sqrt{d}}\sum^{d-1}_{j=0}\exp\left[i \frac{2\pi}{d}j(v+\alpha_a)\right]|j\rangle, \label{eqn:mbasisis1}\\
|w\rangle_b^B=&\tfrac{1}{\sqrt{d}}\sum^{d-1}_{j=0}\exp\left[i \frac{2\pi}{d}j(-w+\beta_b)\right]|j\rangle,
\label{eqn:mbasisis2} 
\end{align}
where $v$ and $w$ both run from $0$ to $d-1$ and denote the outcomes of Alice's and Bob's measurements respectively, and the parameters $\alpha_0=0$, $\alpha_1=1/2$, $\beta_0=1/4$, and $\beta_1=-1/4$. 

The measurement bases $\{|v\rangle_a^A\}$ and $\{|w\rangle_a^A\}$ have been shown \cite{PhysRevLett.104.060401,PhysRevA.64.024101} to maximise the violations of inequality (\ref{eq:s}) for the max\-imal\-ly entan\-gled state of two $d$-dim\-ensional sys\-tems given by $|\psi\rangle=\tfrac{1}{\sqrt{d}}\sum^{d-1}_{j=0}|j\rangle_A\otimes|j\rangle_B.$
It turns out that we are able to parametrise these $d$-dimensional measurement basis states with `mode analyser' angles $\theta_A$ and $\theta_B$, and write them in the form 

\begin{align}
|v\rangle_{a}^A&\equiv|\theta_A^a\rangle=\frac{1}{\sqrt{d}}\sum^{\ell=+[{d}/{2}]}_{\ell=-[{d}/{2}]}\exp\left[i \theta_A^a g(\ell)\right]|\ell\rangle,~{\rm and}  \nonumber \\
|w\rangle_{b}^B&\equiv|\theta_B^b\rangle=\frac{1}{\sqrt{d}}\sum^{\ell=+[\tfrac{d}{2}]}_{\ell=-[\tfrac{d}{2}]}\exp\left[i \theta_B^b g(\ell)\right]|\ell\rangle, \label{eqn:analyserAB}
\end{align}
where
\begin{align}
\theta _A^a
  &= \left( {v + {{a}/{ 2}}} \right){{2\pi }/{d}},\nonumber\\
\theta _B^b
  &= \left[ {-w + {{1}/{4}}\left( { - 1} \right)^b } \right]{{2\pi }/ { d}}.
\label{eqnthetadex} 
\end{align}
The function $g(\ell)$ is defined as
\begin{equation}
g(\ell)=\ell+[\tfrac{d}{2}]+(d\emph{ } \rm{mod}\emph{ }2)u(\ell),
\label{eqn:g}
\end{equation}
where $[x]$ is the integer part of $x$, and $\rm{u}(\ell)$ is the discrete unit step function.

Fig.~\ref{fig:bk2bk} shows an example of the experimental data points for the self normalised coincidence rates as function of the relative angle $(\theta_A-\theta_B)$ using $d=11$ (see also Fig.~S1 in the Supplementary Information). 
For a maximally entangled state 
\begin{equation}
\label{eqn:sxasffw}
|\Phi\rangle =\tfrac{1}{\sqrt{d}} \sum^{[d/2]}_{\ell=-[d/2]}h(\ell){\left| \ell \right\rangle _A \otimes \left| {-\ell} \right\rangle _B },
\end{equation} 
where $h(\ell)=1$ for all $\ell$ when $d$ is odd, and $h(\ell\ne0)=1$, $h(0)=1$ when $d$ is even, 
the coincidence rate of detecting one photon in state $|\theta_A\rangle$ and the other in state $|\theta_B\rangle$ is proportional to
\begin{equation}
\label{eqn:coinccurve}
C(\theta _A ,\theta _B )=| \langle\theta_A|\langle\theta_B||\Phi\rangle|^2 \propto  \frac{\cos (d (\theta _A  - \theta _B ))-1}{d^3 [\cos (\theta _A  - \theta _B )-1]}.
\end{equation}

\begin{figure}[t!]
    \centering
\centerline{\includegraphics[width=0.45\textwidth]{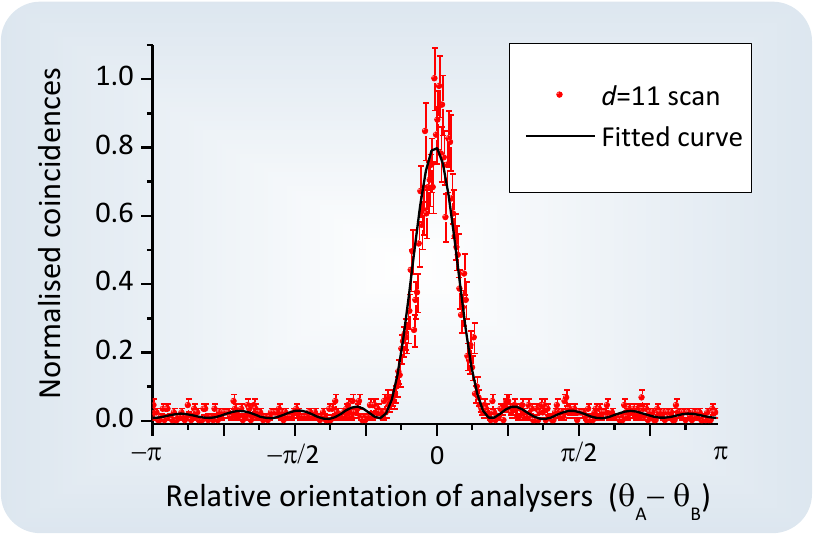}}
 \caption{Coincidence count rates (self normalised) as a functions of relative orientation angles between state analysers $(\theta_A-\theta_B)$. Eqn.~\ref{eqn:coinccurve} for a state with maximal $11$-dimensional entanglement is fitted to the experimental data with the vertical offset and amplitude left as free parameters. Errors were estimated assuming Poisson statistics.}
 \label{fig:bk2bk}
\end{figure}

\begin{figure}[t!]
\centerline{\includegraphics[width=0.45\textwidth]{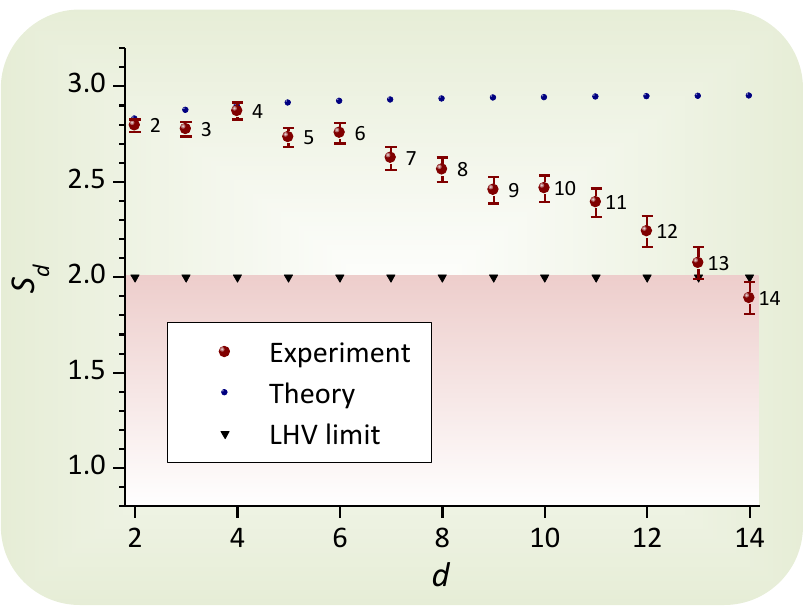}}
\caption{ Experimental Bell-type parameter $S_d$ versus number of dimensions $d$. $S_d>2$ violates local realism for any $d\ge2$.  The plot compares the theoretically predicted violations by a maximally entangled state and the local hidden variable (LHV) limit with the experiments. Violations are observed for up to $d=12$. Errors were estimated assuming Poisson statistics.}
\label{fig:svsdnew}
\end{figure}
 
The key result of our paper is displayed in Fig.~\ref{fig:svsdnew}, which shows a plot of experimental values of parameter $S_d$ as a function of the number of dimensions $d$. The plot compares theoretically predicted violations for a maximally entangled state, the experimental readings and the local hidden variable (LHV) limit. The maximum possible violations (shown in Tab. S1 of the Supplementary Information) are slightly larger than the corresponding violations produced by a maximally entangled state.
 Violations persist up to as much as $d=12$ when entanglement concentration~\cite{PhysRevA.53.2046} is applied. 
We find $S_{11}=2.39\pm0.07$ and $S_{12}=2.24\pm 0.08$, which clearly violate $S_{d} \le 2$ (see also Tab.~S4 of the Supplementary Information). 
In the corresponding experiment using $LG_{p,\ell}$ modes with only $p=0$, violations are obtained up to $d=11$. Without entanglement concentration, we observe violations only up to $d=9$ (please see Fig.~S2 in Sec. SIV of the Supplementary Information).
Above $d \sim 11$, the strength of the signal becomes so low that noise begins to overshadow the quantum correlations. 
In Fig.~\ref{fig:bk2bk}, the theoretical prediction in Eqn.~\ref{eqn:coinccurve} for a state with maximal $11$-dimensional entanglement is fitted to the experimental coincidence data obtained using the mode analyser settings defined in Eq.~\ref{eqn:analyserAB} for $d=11$, with only the vertical offset and amplitude left as free parameters. The observed fringes are seen to closely match those theoretically obtained for a state with maximal 11-dimensional entanglement. 

\begin{figure}[t!]
\centerline{\includegraphics[width=0.45\textwidth]{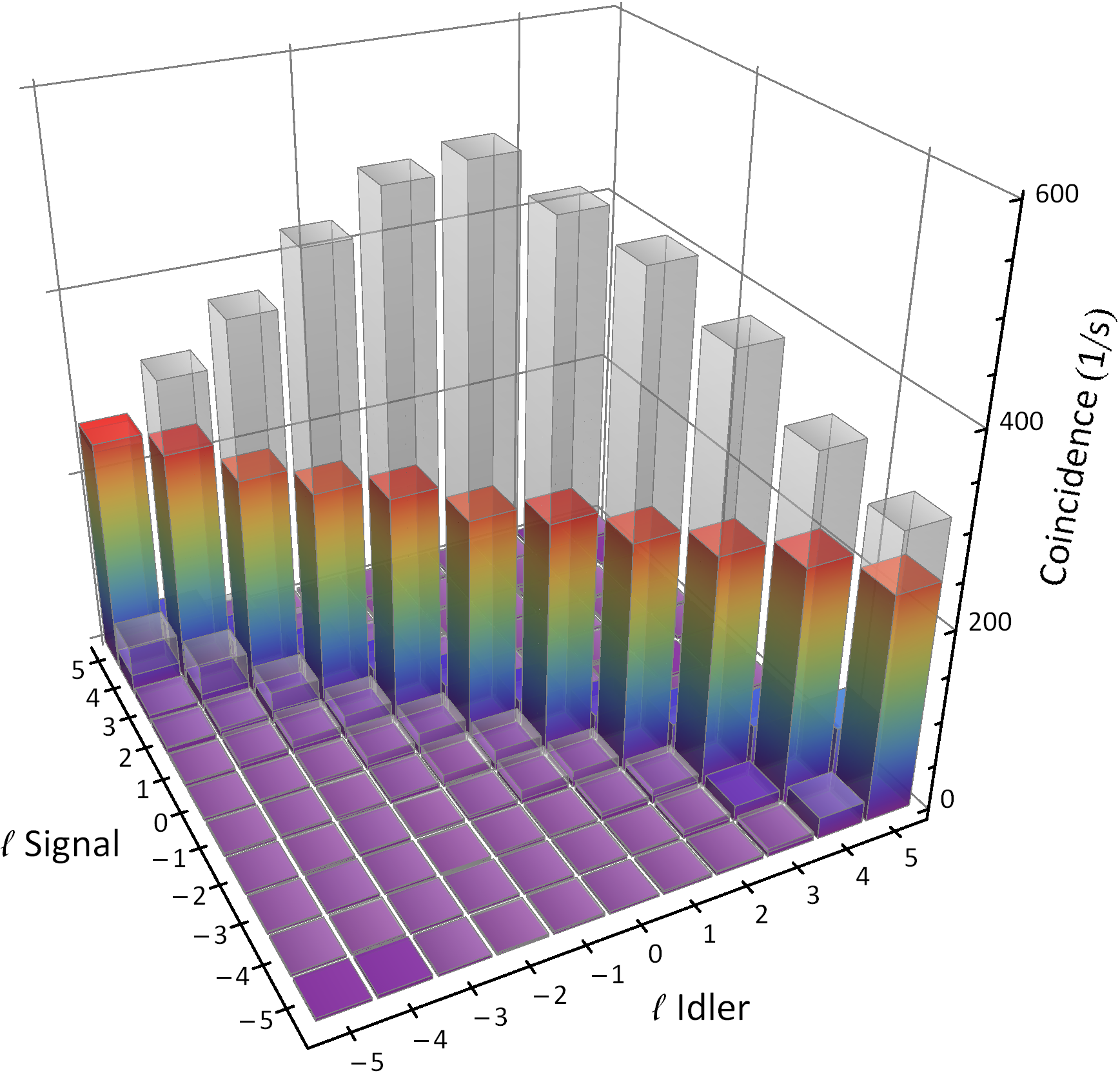}}
\caption{Experimental coincidence rates proportional to the probability of measuring the state $|\ell_s\rangle\otimes|\ell_i\rangle$ with $\ell_s,\ell_i=-5,\ldots,+5$. The coloured and greyed-out bars depict the measurement results with and without the application of Procrustean filtering respectively. The measurement time was $20~\rm{s}$ for each combination of $\ell_s$ and $\ell_i$.}
\label{fig:coincbars}
\end{figure}

The violation of a Bell inequality in $d\times d$ dimensions directly indicates that the measured state was entangled. It remains to determine how many dimensions were involved in the entanglement. Measuring the coincidence probabilities, i.e., of having the joint state  $|\ell_s\rangle\otimes|\ell_i\rangle$ (Fig.~\ref{fig:coincbars}), together with the parameters $S_d$ for different $d$, can be seen as a partial tomography of the SPDC source state. Numerical investigations indicate that a state having the experimentally observed coincidence probabilities and parameters $S_2, S_3, \ldots, S_{11}$ must contain genuine 11-dimensional entanglement. In other words, it is not possible to obtain the observed levels of violation with a state that contains entanglement involving only 10 dimensions or less. Our analysis assumes a special form of the states, based on the coincidence measurement results shown in Fig.~\ref{fig:coincbars}. Further details are given in Sec. SII of the Supplementary Information.

Our results hold much promise for applications requiring entangled qudits in general. As mentioned earlier, progress in the development of quantum repeaters~(see \cite{citeulike:3168245} and references therein) would make quantum key distribution using high-dimensional entangled states~\cite{PhysRevLett.67.661} a possible application. 
Conventional quantum communication will fail for sufficiently large transmission distances because of loss, and quantum repeaters are one possible solution to this problem.
 Although experimental quantum key distribution has been demonstrated with OAM qutrits~\cite{1367-2630-8-5-075}, our findings provide experimental evidence that such protocols could be implemented using photons entangled in OAM in up to $11$ dimensions, resulting in a considerable increase in information coding density.

A possible extension to our work would be to investigate the generation of multi-photon, high-dimensional OAM entanglement. One can conceive of achieving this using a cascade of down-conversion crystals for generating multipartite entangled photons, which has been done for polarisation entangled photons~\cite{citeulike:6916025}. It also appears to be within reach to combine the high-dimensional photon OAM entanglement with entanglement in the polarisation and path degrees of freedom, creating even larger hyper-entangled states~(please see \cite{citeulike:6916025} and references within).

On a more fundamental note, Bell test experiments performed to date have one or both of two main loopholes, namely the locality and detection loopholes.  However, a recent theoretical work reveals that even low dimensional qudits can provide a significant advantage over qubits for closing the detection loophole~\cite{PhysRevLett.104.060401}. In fact, it was found that as much as $38.2\%$ loss can be tolerated using $4$-dimensional entanglement.
Our results raise interesting possibilities regarding the role higher-dimensional entangled qudits could play in closing this loophole. 
We emphasize that neither the detection nor the locality loophole has been closed in our experiments, because the overall efficiency of our experimental setup is $1-2\%$, and the switching time for our measurement devices (SLMs) is of the order of tens of ${\rm ms}$. However, closing these loopholes was not the immediate goal of our experiments. We are instead using the violation of Bell inequalities, up to fair sampling assumptions, as a means of verifying the presence of high-dimensional entanglement, within the framework of quantum mechanics.

In summary, we have been able to experimentally demonstrate violations of Bell-type inequalities generalised to $d$-dimensional systems~\cite{PhysRevLett.88.040404} with up to $d=12$, enough to indicate genuine 11-dimensional entanglement in the orbital angular momentum of signal and idler photons in parametric down-conversion. 
It appears that this could be extended to even higher dimensions by using a brighter source of entangled photons.

\footnotesize
\section*{\bf Methods}
In our experiments, we use computer controlled SLMs (Hamamatsu) operating in reflection mode with a resolution of $600\times600$ pixels. In the detection, the SLMs are prepared in the states defined in Eqns. (\ref{eqn:analyserAB}) respectively. An SLM prepared in a given state transforms a photon in that state to the Gaussian $|\ell=0\rangle$ mode. The reflected photon is then coupled into a single-mode fibre which feeds a single photon detector. Since only the $|\ell=0\rangle$ mode couples into the fibre, a count in the detector indicates a detection of the state in which the SLM was prepared.  The hologram generation algorithm introduced in \cite{1367-2630-7-1-055} is applied to configure the SLMs.

Fig.~\ref{fig:setupoam} shows the schematic diagram of the experimental setup as well as examples of SLM settings used where $d=11$. For the SPDC, we use a pump beam, with $\ell=0$, produced by a frequency tripled, mode-locked Nd-YAG laser with an average output power of $150$ mW at $355$ nm. The collimated laser beam is normally incident on a $3$ mm long BBO crystal cut for type-I collinear phase matching. A $50$:$50$ beam splitter (BS) then separates the co-propagating OAM entangled photons probabilistically into the signal and idler paths. Spectral filters with $10$ nm bandwidth are used to reduce the detection of noise photons. The coincidence resolving time is $10$ ns and an integration time of $20$ s is used for the measurements.

For tests within a $d$-dimensional subspace and for odd $d$, we choose the modes $\ell=-(d-1)/2, \ldots , 0, \ldots, (d-1)/2$ as the computational basis states $|j\rangle$ in Eqns.~\ref{eqn:mbasisis1} and \ref{eqn:mbasisis2}, where $j=0, \ldots d-1$. For even $d$, we use $\ell=-d/2, \ldots, -1, 1, \ldots, d/2$, omitting the $\ell=0$ mode. A projection of the SPDC output state onto a $d$-dimensional subspace results in a non-maximally entangled state due to the limited \emph{spiral bandwidth}~\cite{PhysRevA.68.050301}. To enhance the entanglement, we use the so-called Procrustean method of entanglement concentration~\cite{PhysRevA.53.2046}. This is generally done by means of a filter\-ing tech\-nique which eq\-uali\-ses the mode amp\-li\-tudes, there\-by prob\-abi\-lis\-tical\-ly en\-hancing the entanglement of the two-photon state~\cite{PhysRevLett.92.127903}. This can be achieved by applying local operations to one or both of the signal and idler photons. We choose local operations matched to the spiral bandwidth measurement for our SPDC source  (please see Sec. SIV of the Supplementary Information), so as to obtain a close approximation to a maximally entangled state. The method applied in \cite{PhysRevLett.91.227902} uses lenses for equalising amplitudes in a superposition of three OAM modes. We however use alterations of the diffraction efficiencies of blazed phase gratings in the SLMs to achieve this goal for up to fourteen modes. Fig.~\ref{fig:coincbars} contrasts the results of coincidence measurements with and without Procrustean filtering, with the SLMs in the state  \{$|\ell_A\rangle\otimes|\ell_B\rangle$\} where $\ell_A,\ell_B\in\{-5,\ldots,+5\}$. The disadvantage of the Procrustean method is the associated reduction in the number of detected photons (please see Sec. SIII of the Supplementary Information for further details).\\
\onecolumngrid
\normalsize
{\bf Acknowledgements:}
We acknowledge useful suggestions from S. M. Barnett. This work was funded by the Engineering and Physical Sciences Research Council (EPSRC). A.D. acknowledges funding support from the Scottish Universities Physics Alliance (SUPA).


\section*{}


\renewcommand{\thefigure}{S\arabic{figure}}
\renewcommand{\theequation}{S\arabic{equation}}
\renewcommand{\thesection}{S\Roman{section}}
\renewcommand{\thesubsection}{s\Roman{section}}
\renewcommand{\thetable}{S\arabic{table}}

\setcounter{figure}{0}
\setcounter{equation}{0}
\setcounter{section}{0}

\newpage

\onecolumngrid 
\centering
{\bf \Large Supplementary Information}
\flushleft
\onecolumngrid

\section{Bell operator}
\label{sec:Bellparam}

The expression for the generalised Bell-type parameter derived in~\cite{PhysRevLett.88.040404} can be written as
\small
\begin{align}
\label{eqn:genbell}
S_d = \sum^{[d/2]-1}_{k=0}&\left(1-\tfrac{2k}{d-1}\right)\text{\{}+[ P({A_0=B_0+k})+P({B_0=A_1+k+1})+P(A_1=B_1+k)+P(B_1=A_0+k)]\nonumber\\
&- [ P(A_0=B_0-k-1)+P(B_0=A_1-k)+P(A_1=B_1-k-1)+P(B_1=A_0-k-1)]\text{\}}.
\end{align}
\normalsize
Here, $d$ is the number of dimensions, and \change{$S_d$ is the} Bell parameter, denoted as $I_d$ in~\cite{PhysRevLett.88.040404}.  \change{The measurement outcomes $A_i, B_i \in \{0,\ldots,d-1\}$.}  $P({A_a=B_b})$ denotes the probability that \change{the outcome $A_a$} of Alice's measurement is the same \change{the outcome $B_b$} of Bob's measurement,  for the respective \change{detector settings $a,b\in\{0,1\}$},  (denoted in~\cite{PhysRevLett.88.040404} as $a,b\in\{1,2\}$). That is,
\begin{equation}
\label{eqn:prob1}
P(A_a=B_b)=\sum^{d-1}_{j=0}P(A_a=j,B_b=j).
\end{equation}
\change{In a similar way,}
\begin{equation}
\label{eqn:prob2}
P(A_a=B_b+k)=\sum^{d-1}_{j=0} P\left[A_a=j,B_b= (j + k)\bmod d\right]
\end{equation}
is the probability that \change{the outcome $B_b$ differs from $A_a$ by $k$, modulo $d$.} 
We also have
\begin{equation}
\label{eqn:prob3}
P(B_b=A_a+k)=\sum^{d-1}_{j=0} P\left[A_a=(j + k)\bmod d,B_b=j\right].
\end{equation}
Each joint probability is obtained from the photon coincidence count rate $C(A_a=i,B_b=j)$ divided by the \change{sum $C_T(a,b)$ of all coincidence rates} for a given combination of detector settings for Alice and Bob, $C_T(a,b)=\sum_{i',j'=0}^{d-1}C(A_a=i',B_b=j')$.

\change{The $d$ basis states correspond to orbital angular momentum (OAM) eigenstates, with an OAM of $\ell\hbar$ per photon. For odd $d$, $\ell$ runs from $-(d-1)/2$ to $(d-1)/2$, and for even $d$, $\ell$ runs from $-d/2$ to $+d/2$, but without the $\ell=0$ state. These basis states are denoted as $|j\rangle$, where $j=0,\ldots ,d-1$. 
For the signal photon state and odd $d$, we choose $j=\ell+(d-1)/2$, and for even $d$, similarly, $j=0$ corresponds to $\ell=d/2$, $j=1$ corresponds to $\ell=d/2+1$, and so on, until $j=d-1$ which corresponds to $\ell=d/2$ ($\ell=0$ is missing for even $d$). For the idler photon the ordering is the opposite, with low values of $j$ corresponding to high values of $\ell$. That is, $j=-\ell+(d-1)/2$ for odd $d$, and for even $d$, $j=0$ corresponds to $\ell=d/2$ and so on, until $j=d-1$ which corresponds to $\ell=-d/2$ (again without the $\ell=0$ state). In short, a state $|\ell\rangle\otimes |-\ell\rangle=|j\rangle\otimes|j\rangle=|j,j\rangle$, with the connection between $\ell$ and $j$ as above.
}

As mentioned in the main paper, the Bell-type parameter $S_d$ can be written as the expectation value of a Bell operator $\hat{S}_d$, that is, 
\begin{equation}
S_d(\change{\hat\rho})=Tr(\hat{S}_d\change{\hat\rho}).
\nonumber
\end{equation}
Let us define operators $\hat{P}(A_a=B_b+k)$ as
\begin{equation}
\label{eqn:prob1o}
\hat{P}(A_a=B_b+k)=\sum^{d-1}_{r=0} |r\rangle_a^A|(r + k)\bmod d\rangle_{b}^B{}_{\hskip 1pt b}^B\hskip -1pt\langle (r + k)\bmod d|{}_{\hskip 1pt a}^{A}\hskip-1pt \langle r|,
\end{equation}
and operators $\hat{P}(B_b=A_a+k)$ as
\begin{equation}
\label{eqn:prob2o}
\hat{P}(B_b=A_a+k)=\sum^{d-1}_{r=0} |(r + k)\bmod d\rangle_{a}^A|r\rangle_{b}^B{}_{\hskip 1pt b}^B\hskip -1pt \langle r|{}_{\hskip 1pt a}^A\hskip -1pt \langle (r + k)\bmod d|,
\end{equation}
where the \change{measurement basis} states $|v\rangle_{a}^A$, $|w\rangle_{b}^B$ \change{for measurement settings $a$ and  $b$ and $v, w=0, \ldots , d-1$} are as defined in Eqns. $2$ to $4$ in the main paper.

From Eqn.~\ref{eqn:genbell}, the generalised Bell operator can then be written as
\small
\begin{align}
\label{eqn:genbellop}
\hat{S}_d = \sum^{[d/2]-1}_{k=0}&\left(1-\tfrac{2k}{d-1}\right)\text{\{}+[ \hat{P}({A_0=B_0+k})+\hat{P}({B_0=A_1+k+1}) +\hat{P}(A_1=B_1+k)+\hat{P}(B_1=A_0+k)] \nonumber\\
&- [ \hat{P}(A_0=B_0-k-1)+\hat{P}(B_0=A_1-k)+\hat{P}(A_1=B_1-k-1+\hat{P}(B_1=A_0-k-1)]\text{\}}.
\end{align}
\normalsize
For example, the Bell operators for $d=2$ and $d=3$ are 
\begin{equation}
\hat{S}_2=\left(
\begin{array}{cccc}
 0 & 0 & 0 & 2 \sqrt{2} \\
 0 & 0 & 0 & 0 \\
 0 & 0 & 0 & 0 \\
 2 \sqrt{2} & 0 & 0 & 0
\end{array}
\right),~~~
\hat{S}_3=\left(
\begin{array}{ccccccccc}
 0 & 0 & 0 & 0 & \frac{2}{\sqrt{3}} & 0 & 0 & 0 & 2 \\
 0 & 0 & 0 & 0 & 0 & \frac{2}{\sqrt{3}} & 0 & 0 & 0 \\
 0 & 0 & 0 & 0 & 0 & 0 & 0 & 0 & 0 \\
 0 & 0 & 0 & 0 & 0 & 0 & 0 & \frac{2}{\sqrt{3}} & 0 \\
 \frac{2}{\sqrt{3}} & 0 & 0 & 0 & 0 & 0 & 0 & 0 & \frac{2}{\sqrt{3}} \\
 0 & \frac{2}{\sqrt{3}} & 0 & 0 & 0 & 0 & 0 & 0 & 0 \\
 0 & 0 & 0 & 0 & 0 & 0 & 0 & 0 & 0 \\
 0 & 0 & 0 & \frac{2}{\sqrt{3}} & 0 & 0 & 0 & 0 & 0 \\
 2 & 0 & 0 & 0 & \frac{2}{\sqrt{3}} & 0 & 0 & 0 & 0
\end{array}
\right),
\end{equation}
\change{where the basis states are ordered as $|0,0\rangle$, $|0,1\rangle$, $|0,2\rangle$, \ldots, $|0,d-1\rangle$, $|1,0\rangle$, $|1,1\rangle$,  \ldots, $|1,d-1\rangle$, $|2,0\rangle$, $\ldots$, $|d-1,d-1\rangle$.}

It is straightforward to check that \change{ all diagonal elements of $\hat{S}_d$ are equal to zero in the OAM basis $|j,k\rangle$}, as all \change{diagonal elements of the operators $\hat{P}(A_a=B_b+k)$ and the operators $\hat{P}(B_b=A_a+k)$ are equal to $1/d^2$}. \change{Also, we note that all elements $S_{jk}$ of the Bell operators, written in the OAM basis, are nonnegative. This in fact holds for all Bell operators until $d=14$, and we suspect that it holds generally, although it is not necessary for the purposes of this paper to prove it for the general case.} 

\change{Clearly, the maximal value of the Bell parameter, that is, the maximal Bell violation, is the largest eigenvalue of the Bell operator, and this largest violation is obtained for the corresponding eigenstate. Interestingly, this eigenstate is not in general equal to a maximally entangled state.}
Table~\ref{tab:belleignvals} lists the Bell parameter values \change{theoretically} obtained for the maximally entangled state $|\psi\rangle\change{=(1/\sqrt{d})\sum_{j=0}^{d-1}|j,j\rangle}$, together with the maximal values of the Bell parameter, \change{which, as already stated above, are equal to the} largest eigenvalue of the respective Bell operator~\cite{PhysRevA.65.052325}.
\change{For any $\hat\rho$, the corresponding Bell violation is a linear combination of the eigenvalues of $\hat S_d$, since
\begin{equation}
S_d=Tr(\hat\rho \hat S_d)=\sum_k s_k \langle s_k|\hat\rho|s_k\rangle,
\end{equation} 
where $s_k$ and $|s_k\rangle$ are the eigenvalues and eigenvectors of $\hat S_d$, and therefore $\sum_k \langle s_k|\hat\rho|s_k\rangle={\rm Tr}\hat \rho=1$.
The five largest eigenvalues $s_k$ of $\hat{S}_{11}$ are shown in Table~\ref{tab:belleign11v}, \change{along with} the \change{form of the corresponding eigenstates $|s_k\rangle$.}
}

\section{Evidence for high-dimensional entanglement}
\label{ref:proof}
Violation of a Bell inequality indicates that the measured state was entangled, but does not directly give information of how many dimensions were participating in the entanglement. We will now analyse the results of the coincidence measurements of the SPDC output, shown in Fig.~4 of the main paper, together with the obtained Bell violations for different $d$, and show that the high levels of violations of the tested Bell inequalities indicate that the SPDC state is indeed a high-dimensional entangled state. The problem of finding a lower bound on the dimensionality required to produce certain quantum correlations has been studied by Acin \emph{et al.} in~\cite{PhysRevLett.100.210503}, where they present examples of correlations that require measurements on quantum systems of dimension greater than two for their generation. Although, the depth of analysis presented there would require immense computational efforts for 11 dimensions, additional experimental results and considerations below allow us to simplify our analysis.

Assuming conservation of OAM, meaning that $j$ is always the same for signal and idler photons, both the filtered and unfiltered SPDC source states can be written in the form
\begin{equation}
\hat\rho=\sum_{j,k=0}^{d-1} c_{jk}|j,j\rangle\langle k,k|.
\label{eq:spdcrho}
\end{equation}
As seen from \red{Fig.~4} in the main paper, for high $|\ell |$, there are nonzero coincidence counts also when $j$ for the signal and idler photons differ by $\pm 1$. This occurs for approximately $8\%$ of the total count rate. Conservation of angular momentum in parametric down-conversion is, however, a theoretically and experimentally well-documented phenomenon~(please see Ref. [9] of the main paper, and references within). One is therefore justified in assuming a state of the form in Eq.~\eqref{eq:spdcrho}, and that any coincidence counts indicating unequal $j$ for signal and idler photons are due to imperfect mode selectivity in the measurement, which has greater effect for higher $|\ell |$.

A state of the form in Eqn. \eqref{eq:spdcrho}, which in addition has only at most $(d-1)$-dimensional entanglement, can be written as a mixture of pure states \red{$|\psi_m\rangle=\sum_{j=0}^{d-1} a_{mj}|j,j\rangle$}, where for each $|\psi_m\rangle$, it holds that $a_{mj}=0$ for at least one $j$. By grouping terms with different $|\psi_m\rangle$ together, we can further write a state with at most $(d-1)$-dimensional entanglement as a mixture of no more than $d$ mixed states, $\hat\rho = \sum_{n=0}^{d-1} r_n \hat\rho_n$, where each $\hat\rho_n$ is a mixture of states $|\psi_m\rangle$ for which $a_{mn}=0$.  That is,  if $\hat\rho_n=\sum_{j,k=0}^{d-1} c_{n,jk}|j,j\rangle\langle k,k|$, then $c_{n,jk}=0$ whenever one or both of $j,k$ are equal to $n$.

By varying the $r_n$ and the $c_{n,jk}$ in
\begin{equation}
\hat\rho=\sum_{n=0}^{d-1} r_n\hat\rho_n=\sum_{n=0}^{d-1} r_n \sum_{j,k = 0 }^{d-1} c_{n,jk} | j,j\rangle\langle k,k |,
\label{eq:rhoansatz}
\end{equation}
one may now investigate what level of Bell violations may be obtained for a state with at most $(d-1)$-dimensional entanglement. \red{Clearly, the maximum violation for such a state is attained when it is pure, \emph{if there are no constraints in the maximisation}. When constraints are included, it is possible in general that there is no pure state which satisfies all of them, so that the maximum is obtained with a mixed state. For example, three constraints on the Bloch vector of a 2-dimensional state may force it to be mixed}. For $d\geq 4$ it is in fact enough to consider pure states $\hat\rho_n=|\psi_n\rangle\langle\psi_n|$. This follows since, for any mixed $\hat \rho_n$, there is a corresponding pure state with the same diagonal components $c_{n,jj}$, and the same Bell violations for $S_2, S_3,\ldots , S_d$. To see this, note that in addition to the $d-1$ diagonal components of $\hat\rho_n$, we constrain the values of $d-1$ Bell parameters, and this constitutes in total $2d-2$ constraints on $\hat\rho_n$, which has $(d-1)^2-1$ free parameters. These parameters can be thought of as the diagonal components of $\hat\rho_n$, plus the real and imaginary parts of its off-diagonal elements. We can also think of them as the components of the generalised Bloch vector for $\hat\rho_n$. If the constraints leave at least one component of the generalised Bloch vector free, then this last component can be chosen large enough for $\hat\rho_n$ to be a pure state. We thus require that $(d-1)^2-1> 2d-2$, which is satisfied for $d\ge 4$. For $d\ge5 $, we can in fact choose the $\hat\rho_n$ not only to be pure but also to have only real matrix elements, since then the number of independent off-diagonal matrix elements in $\hat\rho_n$ is strictly greater than the number of Bell inequalities, that is, $(d-1)(d-2)/2>d-1$ if $d\ge 5$. This considerably simplifies the numerical maximisation, which otherwise would be relatively demanding. 

We focus on the case $d=11$, since the level of violation for $d=12$ is smaller. The corresponding state space is thus spanned by the OAM basis states $|j\rangle\otimes|k\rangle=|j,k\rangle$, with $j,k\in\{0,\ldots,10\}$. For $d=11$, we now use the experimentally obtained diagonal coincidence probabilities $P_{jj}$, with $j=0,\ldots ,10$, and the experimental values of $S_2, S_3, \ldots , S_{10}$ as constraints. All these quantities are allowed to vary within the experimental error bars; errors are estimated assuming Poisson statistics for the photon counting processes. By numerical maximisation using $Mathematica^{\small \textregistered}$, we find that the largest value of $S_{11}$, that can be obtained with a state of the form in Eq. \eqref{eq:rhoansatz}, is $S_{11}=2.14$. As mentioned in the main paper, we measure $S_{11}$ to be $2.39\pm0.07$ using all $p$ modes. Using only $p=0$, we find $2.67\pm0.22$. These and the other experimentally obtained Bell parameters are listed in Tables S3 and S4. In other words, the largest violation $S_{11}$ that can be obtained with a state that has at most 10-dimensional entanglement is smaller than the experimentally measured violation, with more than three standard deviations for all $p$ modes, and with more than two standard deviations for $p=0$. One may therefore conclude, with high confidence, that the SPDC output state does contain 11-dimensional entanglement.
\newpage
\section{Procrustean filtering}
\label{sec:proc}
Let us denote the unfiltered state as $\hat{\rho}_i$.
\change{The Procrustean entanglement concentration we perform can be considered to be a two-outcome generalised measurement (POM or probability operator measure) with measurement operators $\hat O_1^\dagger \hat O_1$ and $\hat O_2^\dagger \hat O_2$. The Hermitian operators $\hat O_1$ and $\hat O_2$
satisfy $\hat O_1^\dagger \hat O_1+\hat O_2^\dagger \hat O_2={\bf 1}$ in the relevant Hilbert space. Furthermore, the procedure can be performed using only local operations on the signal and idler beams.
The filtered state is obtained conditioned on outcome 1, and is given by
\begin{equation}
\hat{\rho}_f=\frac{\hat{O}_1^\dag\hat{\rho}_i\hat{O}_1}{{\rm Tr}(\hat{O}_1^\dag\hat{\rho}_i\hat{O}_1)}.
\end{equation}
If outcome 2 is obtained, then the filtering has failed, and in our case, we can think of the photon as having been absorbed, leaving the vacuum state, $\hat{O}_2^\dag\hat{\rho}_i\hat{O}_2
/{\rm Tr}[\hat{O}_2^\dag\hat{\rho}_i\hat{O}_2)]=|0\rangle\langle 0|$.}

Using the filtered state $\hat{\rho}_f$, the parameter $S_d$ is
\begin{equation}
S_d= Tr(\hat{\rho}_f \hat{S}_d) = {\rm Tr}\left[\frac{\hat{O}^\dag\hat{\rho}_i\hat{O}}{{\rm Tr}(\hat{O}^\dag\hat{\rho}_i\hat{O})} \hat{S}_d\right].
\end{equation}

The Procrustean filter we applied for all the experiments was by means of local operations of the form $\hat{O_1}=\hat{O}_A\otimes\hat{O}_B$, where $\hat{O}_A=\hat{O}_B\equiv\hat{o}$ is approximately a diagonal matrix. \change{For $d$=11, $\hat o$ has diagonal elements (1.00, 0.97, 0.94, 0.92, 0.91, 0.90, 0.91, 0.92, 0.93, 0.95, 0.97).}

Since the purpose of our experiment is to demonstrate entanglement, within the framework of quantum mechanics (i.e., assuming that we are testing some quantum mechanical state for entanglement), it is important to note that the local filtering can produce neither the appearance of entanglement nor the violation of a Bell inequality from a separable state. 
Any separable state can be written as 
\change{$\hat{\rho}=\sum_m p_m \hat{\rho}_{m,A}\otimes\hat{\rho}_{m,B}.$}
After filtering, we have
\change{$\hat\rho_f\propto\hat{O_1}^\dag\hat{\rho}\hat{O_1}
=\sum_m p_m (\hat{o}\hat{\rho}_{m,A}\hat{o})\otimes(\hat{o}\hat{\rho}_{m,B}\hat{o})$},
which is just another separable state.
No separable state can produce a violation of any of the inequalities $S_d\le2$. 

\section{Supplementary Experimental Results}
\label{sec:suppresults}

{\bf Spiral bandwidth.} The spiral bandwidth is an important factor affecting the amount of entanglement of the OAM correlated photons for a given selected subspace of the OAM Hilbert space. The square of a Lorentzian function
\begin{equation}
\label{eqn:lortz}
f(\ell,\gamma)=A {\gamma}/{\left(\gamma^2+\ell^2\right)},
\end{equation}
gives a good fit to the experimentally measured coincidence rates, which are proportional to $|c_\ell|^2$ and presented in Fig.~4 of the main paper. The parameter $\gamma$ specifies the half-width at half-maximum (HWHM) of $f(\ell,\gamma)$,  and $A$ is a normalisation constant. We identify $\gamma$ with the effective quantum spiral bandwidth. For our source, we obtain $\gamma$ to be $7.58$. As mentioned in the main paper, the finiteness of the spiral bandwidth causes a projection of the SPDC output state onto a $d$-dimensional subspace to result in a non-maximally entangled state~\cite{PhysRevA.68.050301}. 

{\bf $S_d$ versus $d$ with $LG_{p,\ell}$ modes.}
Table~\ref{tab:sbelld1} shows the values plotted in Fig. 3 of the main paper, using all radial modes (all $p$). This means that the OAM (denoted by $\ell$) states of the photons are used irrespective of their radial state (denoted by $p$), which is another degree of freedom of Laguerre Gaussian beams. \change{The corresponding theoretical maximal violations are shown in Table S1. }

{\bf $S_d$ versus $d$ using $LG_{p,\ell}$ modes without Procrustean filtering.}
Figure~\ref{fig:svsdnewnew}~(a) shows the experimental results (also displayed in Table~\ref{tab:sbelldsupp2}) obtained without Procrustean filtering.

{\bf $S_d$ versus $d$ with $LG_{p=0,\ell}$ modes.}
The following results were obtained with the radial index $p$ of the detected beams restricted to only $p=0$. In this case, the state $|\ell\rangle$ means $|p=0,\ell\rangle$. The data in Table~\ref{tab:sbelldsupp} is plotted in Figure~\ref{fig:svsdnewnew}~(b). \change{This figure} compares theoretically predicted violations for a maximally entangled state, the experimental readings and the local hidden variable (LHV) limit. Again, the corresponding theoretical maximal violations are shown in Table S1. Using only modes with radial index $p$=0, we observe  increased error bars and violations only up to $d=11$, due to the reduced count rates resulting from the detection of only $p=0$ states.

{\bf Coincidence curves obtained using $LG_{p=0,\ell}$ modes.} 
 As defined in the main paper, the measurement states in terms of `mode analyser' angles $\theta_A$ and $\theta_B$, are 
\begin{equation}
\label{eqn:analyserA}
|\theta_A\rangle=\frac{1}{\sqrt{d}}\sum^{\ell=+[{d}/{2}]}_{\ell=-[{d}/{2}]}\exp\left[i \theta_Ag(\ell)\right]|\ell\rangle,~~~~~{\rm and}~~~~~
|\theta_B\rangle=\frac{1}{\sqrt{d}}\sum^{\ell=+[\tfrac{d}{2}]}_{\ell=-[\tfrac{d}{2}]}\exp\left[i \theta_Bg(\ell)\right]|\ell\rangle.
\end{equation}
 Fig.~\ref{fig:bk2bk}(a) and Fig.~\ref{fig:bk2bk}(b) suggest reasonable agreement between the experimental coincidence measurements and the theoretical predictions for maximally entangled states.
\begin{figure*}[t!]
    \centering
\includegraphics[width=1\textwidth]{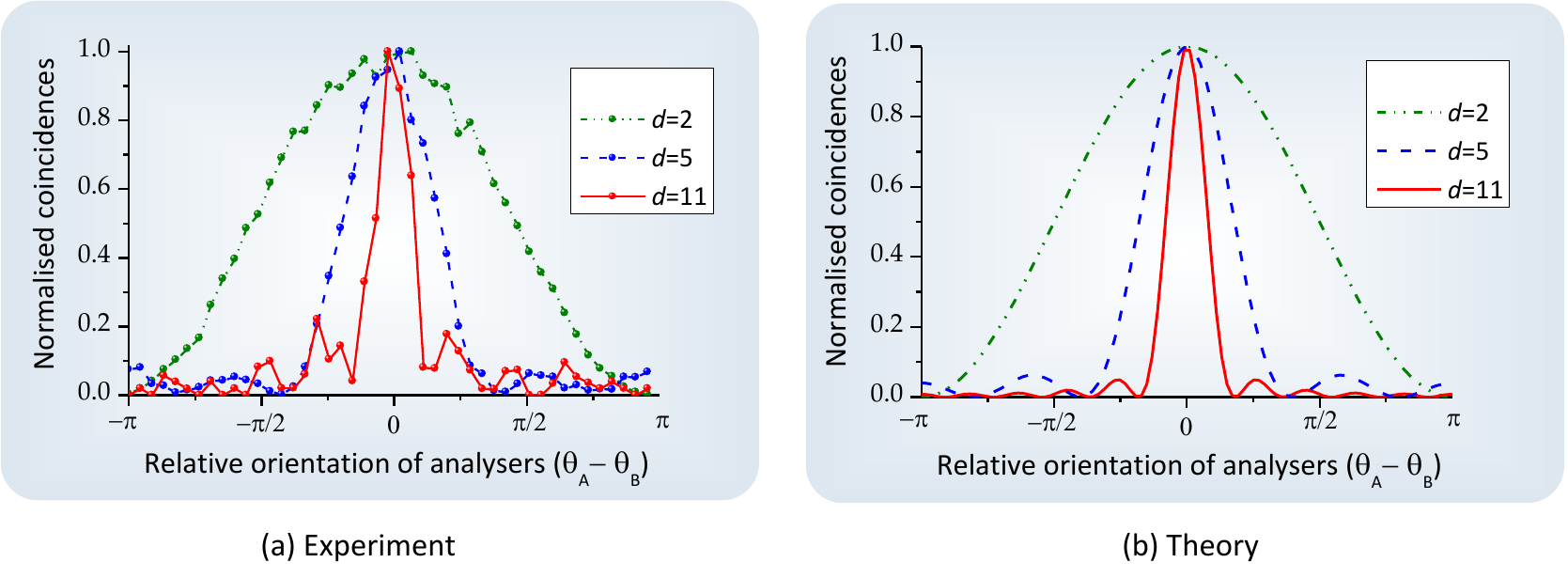}
 \caption{Coincidence count rates (self normalised) as a functions of relative orientation angles between state analysers $(\theta_A-\theta_B)$. \change{The figures depict (a) experimental coincidence curves and (b) theoretical prediction for maximally entangled states of two $d$-dimensional systems with mode analyser settings defined in Eq.~\ref{eqn:analyserA} using $d=2,5,$ and $11$, as examples. Only modes with $p=0$ are used here. The observed fringes are typical of genuine 2-, 5-, and 11-dimensional entanglement, respectively}.}
 \label{fig:bk2bk}
\end{figure*} 
\vspace{30pt}
\begin{figure*}[t]
    \centering
\includegraphics[width=1\textwidth]{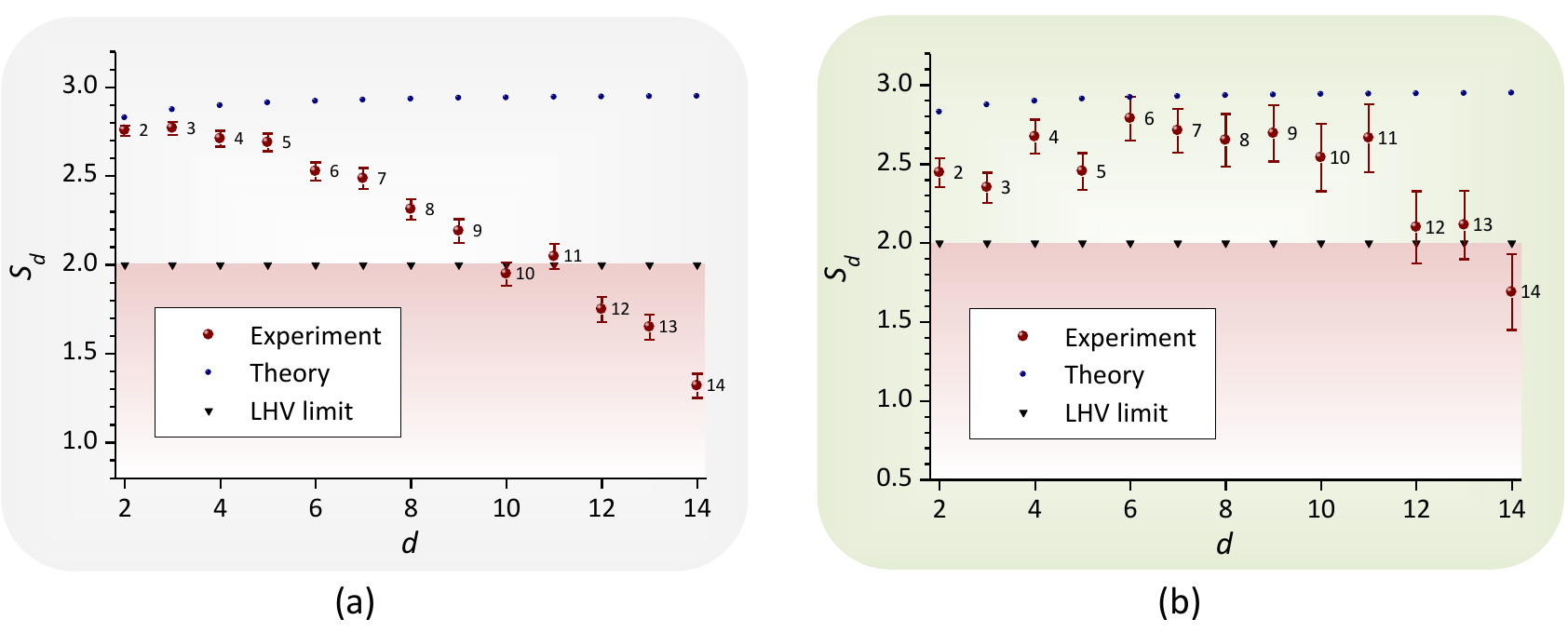}
 \caption{ Supplementary experimental results for $S_d$ versus number of dimensions $d$ \change{using (a) all $p$ modes and no procrustean filtering and (b) only $p=0$ modes with procrustean filtering. $S_d>2$ violates local realism for any $d\ge2$. The plots compares the theoretically predicted violations by a maximally entangled state and the local hidden variable (LHV) limit with the experiments. Violations are observed using up to $d=11$ dimensions for each qudit. Errors were calculated assuming Poisson statistics for the photon counting processes. The measurement time was $20$ s per point}.}
 \label{fig:svsdnewnew}
\end{figure*}


\begin{table*}[h!]
\begin{tabular}{c| c| c}
\hline \hline
~~No. of Dimensions ~~$d$	~~&~~Violation for $|\psi\rangle$~~&~~ Maximum possible violation~~	\\\hline
2 & 2.8284 & 2.8284\\
3 & 2.8729 & 2.9149\\
4 & 2.8962 & 2.9727\\
5 & 2.9105 & 3.0157\\
6 & 2.9202 & 3.0497\\
7 & 2.9272 & 3.0776\\
8 & 2.9324 & 3.1013\\
9 & 2.9365 & 3.1217\\
10 & 2.9398 & 3.1396\\
11 & 2.9425 & 3.1555\\
12 & 2.9448 & 3.1698\\
13 & 2.9467 & 3.1827\\
14 & 2.9483 & 3.1946\\
\hline \hline
\end{tabular}
\caption[Violation of the inequality $S_d\le2$ for two $d$-dimensional qudits, up to $d=14$.] {Violation of the inequality $S_d\le2$ for two $d$-dimensional qudits, up to $d=14$. This Table shows theoretically predicted values obtained for the maximally entangled state $|\psi\rangle$ (depicted in Fig.~$3$ of the main paper), and the maximum possible violation corresponding to the largest eigenvalue of the Bell operator $\hat{S}_d$.}
\label{tab:belleignvals}
\vspace{5pt}
\begin{tabular}{c| c| c}
\hline \hline
~~Eigenvalue of $\hat{S}_{11}$~~&~~~~~~~~~~~~~~Eigenstate~~~~~~~~~~~~~~&~~~\ldots -dimensional entanglement~~~	\\\hline
3.1555 & $|s_1\rangle\equiv\sum_{n=0}^{10} C_{1,n} |n,n\rangle$ & 11\\
2.4107 & $|s_2\rangle\equiv\sum_{n=0}^{9} C_{2,n} |n,n+1\rangle$ & 10\\
2.4107 & $|s_3\rangle\equiv\sum_{n=0}^{9} C_{3,n} |n+1,n\rangle$ & 10\\
1.9709 & $|s_4\rangle\equiv\sum_{n=0}^{8} C_{4,n} |n,n+2\rangle$ & 9\\
1.9709 & $|s_5\rangle\equiv\sum_{n=0}^{8} C_{5,n} |n+2,n\rangle$ & 9\\
\hline \hline
\end{tabular}
\caption[Five largest eigenvalues of $\hat{S}_{11}$.] {Five largest eigenvalues of $\hat{S}_{11}$ and the form of the corresponding eigenstate.}
\label{tab:belleign11v}
\vspace{20pt}
\begin{tabular}{|c | c |c |c|c | c |c |c|c | c |c |c|c | c |}
\hline \hline
No. of Dimensions $d$	&2	&3	&4&5	&6	& 7	 & 8 	& 9 &	 10 	& 11 	& 12 	& 13 	& 14 \\\hline
Parameter $S_d$	&~2.79~	&~2.78	~&~2.87	&~2.73~	&~2.76~&~2.62~	&~2.56~	&~2.46~	&~2.47~	&~2.39~	&~2.24~	&~2.07~	&~1.89~\\\hline
Standard deviation $\sigma$	&0.03	&0.04	&0.04	&0.05	&0.06	&0.07	&0.07	&0.07	&0.07	&0.07	&0.08	&0.08	&0.08\\\hline\hline
 \end{tabular}
\caption[$S_d$ values for different number of dimensions $d$ in a round of experiments.] {Experimental Bell-type parameter $S_d$ as a function of the number of dimensions $d$. All radial modes (all $p$) are used here, and violations of $S_d\le2$ are observed for up to $d=12$.}
\label{tab:sbelld1}
\end{table*}
\begin{table*}[th]
\begin{tabular}{|c | c |c |c|c | c |c |c|c | c |c |c|c | c |}
\hline \hline
No. of Dimensions $d$	&2	&3	&4&5	&6	& 7	 & 8 	& 9 &	 10 	& 11 	& 12 	& 13 	& 14 \\\hline
Parameter $S_d$	&~2.45~	&~2.4	~&~2.67	&~2.46~	&~2.79~&~2.71~	&~2.65~	&~2.7~	&~2.54~	&~2.67~	&~2.1~	&~2.11~	&~1.69~\\\hline
Standard deviation $\sigma$	&0.09	&0.1	&0.11	&0.12	&0.14	&0.14	&0.16	&0.2	&0.21	&0.22	&0.2	&0.22	&0.24\\\hline\hline
 \end{tabular}
\caption[$S_d$ values for different number of dimensions $d$ in a round of experiments.] {Experimental Bell-type parameter $S_d$ as a function of the number of dimensions $d$. Only OAM modes with $p=0$ are used here, and violations of $S_d\le2$ are observed for up to $d=11$.}\label{tab:sbelldsupp}
\end{table*}
\begin{table*}[h!]
\begin{tabular}{|c | c |c |c|c | c |c |c|c | c |c |c|c | c |}
\hline \hline
No. of Dimensions $d$	&2	&3	&4&5	&6	& 7	 & 8 	& 9 &	 10 	& 11 	& 12 	& 13 	& 14 \\\hline
Parameter $S_d$	&~2.76~	&~2.77	~&~2.71	&~2.69~	&~2.53~&~2.49~	&~2.31~	&~2.19~	&~1.95~	&~2.05~	&~1.75~	&~1.65~	&~1.32~\\\hline
Standard deviation $\sigma$	&0.03	&0.04	&0.04	&0.05	&0.05	&0.06	&0.06	&0.07	&0.07	&0.07	&0.07	&0.07	&0.07\\\hline\hline
 \end{tabular}
\caption[$S_d$ values for different number of dimensions $d$ in a round of experiments.] {Experimental Bell-type parameter $S_d$ as a function of the number of dimensions $d$ without entanglement concentration.  Violations of $S_d\le2$ are observed for up to $d=9$. All $p$ modes are used here.}\label{tab:sbelldsupp2}
\end{table*}

\subsection*{}


\begin{thebibliography}{10}
\onecolumngrid
\bibitem{PhysRev.47.777}
Einstein, A., Podolsky, B., \& Rosen, N.
\newblock Can Quantum-Mechanical Description of Physical Reality Be Considered
  Complete?
\newblock {\em Phys. Rev.}{ \bf 47}, 777--780 (1935).

\bibitem{schro1935}
Schr\"odinger, E.
\newblock Die gegenw\"artige Situation in der Quantenmechanik.
\newblock {\em Naturwissenschaften}{ \bf 23}, 807--812; 823--828; 844--849
  (1935).

\bibitem{Nielsen2000}
Nielsen, M.~A. \& Chuang, I.~L.
\newblock {\em Quantum Computation and Quantum Information}.
\newblock Number ISBN 0-521-63503-9. Cambridge University Press, New York,
  (2000).

\bibitem{PhysRevA.69.032313}
Durt, T., Kaszlikowski, D., Chen, J.-L., \& Kwek, L.~C.
\newblock Security of quantum key distributions with entangled qudits.
\newblock {\em Phys. Rev. A}{ \bf 69}, 032313 (2004).

\bibitem{PhysRevLett.104.060401}
V\'ertesi, T., Pironio, S., \& Brunner, N.
\newblock Closing the Detection Loophole in Bell Experiments Using Qudits.
\newblock {\em Phys. Rev. Lett.}{ \bf 104}, 060401 (2010).

\bibitem{PhysRevLett.88.040404}
Collins, D., Gisin, N.-l., Lin\-den, N., Mas\-sar, S., \& Popescu, S.
\newblock Bell Inequa\-lities for Arbitra\-rily High\-Dimen\-sional Systems.
\newblock {\em Phys. Rev. Lett.}{ \bf 88}, 040404 (2002).

\bibitem{PhysRevA.45.8185}
Allen, L., Beijersbergen, M.~W., Spreeuw, R. J.~C., \& Woerdman, J.~P.
\newblock Orbital angular momentum of light and the transformation of
  Laguerre-Gaussian laser modes.
\newblock {\em Phys. Rev. A}{ \bf 45}, 8185--8189 (1992).

\bibitem{PhysRevLett.75.4337}
Kwiat, P.~G., \emph{et~al.}
\newblock New High-Intensity Source of Polarization-Entangled Photon Pairs.
\newblock {\em Phys. Rev. Lett.}{ \bf 75}, 4337--4341 (1995).

\bibitem{PhysRevA.69.023811}
Walborn, S.~P., de~Oliveira, A.~N., Thebaldi, R.~S., \& Monken, C.~H.
\newblock Entanglement and conservation of orbital angular momentum in
  spontaneous parametric down-conversion.
\newblock {\em Phys. Rev. A}{ \bf 69}, 023811 (2004).

\bibitem{PhysRevLett.70.1895}
Bennett, C.~H., \emph{et~al.}
\newblock Teleporting an unknown quantum state via dual classical and
  Einstein-Podolsky-Rosen channels.
\newblock {\em Phys. Rev. Lett.}{ \bf 70}, 1895--1899 (1993).

\bibitem{doi:827398}
You-Bang, Z., Qun-Yong, Z., Yu-Wu, W., \& Peng-Cheng, M.
\newblock Schemes for Teleportation of an Unknown Single-Qubit Quantum State by
  Using an Arbitrary High-Dimensional Entangled State.
\newblock {\em Chinese Physics Letters}{ \bf 27}, 10307--10310 (2010).

\bibitem{PhysRevLett.69.2881}
Bennett, C.~H. \& Wiesner, S.~J.
\newblock Communication via one- and two-particle operators on
  Einstein-Podolsky-Rosen states.
\newblock {\em Phys. Rev. Lett.}{ \bf 69}, 2881--2884 (1992).

\bibitem{PhysRevLett.67.661}
Ekert, A.~K.
\newblock Quantum cryptography based on Bell's theorem.
\newblock {\em Phys. Rev. Lett.}{ \bf 67}, 661--663 (1991).

\bibitem{PhysRevA.59.1829}
Hillery, M., Bu\ifmmode~\check{z}\else \v{z}\fi{}ek, V., \& Berthiaume, A.
\newblock Quantum secret sharing.
\newblock {\em Phys. Rev. A}{ \bf 59}, 1829--1834 (1999).

\bibitem{PhysRevA.68.022312}
Raussendorf, R., Browne, D.~E., \& Briegel, H.~J.
\newblock Measurement-based quantum computation on cluster states.
\newblock {\em Phys. Rev. A}{ \bf 68}, 022312 (2003).

\bibitem{citeulike:3168245}
Yuan, Z.-S., \emph{et~al.}
\newblock Experimental demonstration of a BDCZ quantum repeater node.
\newblock {\em Nature}{ \bf 454}, 1098--1101 (2008).

\bibitem{citeulike:6916025}
Gao, W.-B., \emph{et~al.}
\newblock {Experimental demonstration of a hyper-entangled ten-qubit
  Schr\"{o}dinger cat state}.
\newblock {\em Nature Physics}{ \bf 6}, 331--335 (2010).

\bibitem{PhysRevLett.89.240401}
Vaziri, A., Weihs, G., \& Zeilinger, A.
\newblock Experimental Two-Photon, Three-Dimensional Entanglement for Quantum
  Communication.
\newblock {\em Phys. Rev. Lett.}{ \bf 89}, 240401 (2002).

\bibitem{PhysRevLett.49.91}
Aspect, A., Grangier, P., \& Roger, G.
\newblock Experimental Realization of Einstein-Podolsky-Rosen-Bohm
  Gedankenexperiment: A New Violation of Bell's Inequalities.
\newblock {\em Phys. Rev. Lett.}{ \bf 49}, 91--94 (1982).

\bibitem{Jleach2009}
Le\-ach, J., \emph{et~al.}
\newblock Violation of a Bell inequality in two-dimensional orbital angular
  momentum state-spaces.
\newblock {\em Optics Express}{ \bf 17}, 8287--8293 (2009).

\bibitem{PhysRevA.81.043844}
Jack, B., \emph{et~al.}
\newblock Entanglement of arbitrary superpositions of modes within
  two-dimensional orbital angular momentum state spaces.
\newblock {\em Phys. Rev. A}{ \bf 81}, 043844 (2010).

\bibitem{1367-2630-11-10-103024}
Jack, B., \emph{et~al.}
\newblock Precise quantum tomography of photon pairs with entangled orbital
  angular momentum.
\newblock {\em New Journal of Physics}{ \bf 11}, 103024 (2009).

\bibitem{PhysRevLett.101.120502}
Pors, J.~B., \emph{et~al.}
\newblock Shannon Dimensionality of Quantum Channels and Its Application to
  Photon Entanglement.
\newblock {\em Phys. Rev. Lett.}{ \bf 101}, 120502 (2008).

\bibitem{1367-2630-8-5-075}
Gr{\"o}blacher, S., Jennewein, T., Vaziri, A., Weihs, G., \& Zeilinger, A.
\newblock Experimental quantum cryptography with qutrits.
\newblock {\em New Journal of Physics}{ \bf 8}, 75 (2006).

\bibitem{PhysRevA.64.024101}
Durt, T., Kaszlikowski, D., \& \ifmmode~\dot{Z}\else \.{Z}\fi{}ukowski, M.
\newblock Violations of local realism with quantum systems described by
  N-dimensional Hilbert spaces up to $N=16$.
\newblock {\em Phys. Rev. A}{ \bf 64}, 024101 (2001).

\bibitem{PhysRevA.53.2046}
Bennett, C.~H., Bernstein, H.~J., Popescu, S., \& Schumacher, B.
\newblock Concentrating partial entanglement by local operations.
\newblock {\em Phys. Rev. A}{ \bf 53}, 2046--2052 (1996).

\bibitem{1367-2630-7-1-055}
Leach, J., Dennis, M.~R., Courtial, J., \& Padgett, M.~J.
\newblock Vortex knots in light.
\newblock {\em New Journal of Physics}{ \bf 7}, 55 (2005).

\bibitem{PhysRevA.68.050301}
Torres, J.~P., Alexandrescu, A., \& Torner, L.
\newblock Quantum spiral bandwidth of entangled two-photon states.
\newblock {\em Phys. Rev. A}{ \bf 68}, 050301 (2003).

\bibitem{PhysRevLett.92.127903}
Law, C.~K. \& Eberly, J.~H.
\newblock Analysis and Interpretation of High Transverse Entanglement in
  Optical Parametric Down Conversion.
\newblock {\em Phys. Rev. Lett.}{ \bf 92}, 127903 (2004).

\bibitem{PhysRevLett.91.227902}
Vaziri, A., Pan, J.-W., Jennewein, T., Weihs, G., \& Zeilinger, A.
\newblock Concentration of Higher Dimensional Entanglement: Qutrits of Photon
  Orbital Angular Momentum.
\newblock {\em Phys. Rev. Lett.}{ \bf 91}, 227902 (2003).

\end{thebibliography}

\begin{thebibliography}{2}

\bibitem{PhysRevLett.88.040404}
Collins, D., Gisin, N.-l., Lin\-den, N., Mas\-sar, S., \& Popescu, S.
\newblock Bell Inequa\-lities for Arbitra\-rily High\-Dimen\-sional Systems.
\newblock {\em Phys. Rev. Lett.}{ \bf 88}, 040404 (2002).

\bibitem{PhysRevA.65.052325}
Ac\'\i{}n, A., Durt, T., Gisin, N., \& Latorre, J.~I.
\newblock Quantum nonlocality in two three-level systems.
\newblock {\em Phys. Rev. A}{ \bf 65}, 052325 (2002).

\bibitem{PhysRevLett.100.210503}
Brunner, N., \emph{et~al.}
\newblock Testing the Dimension of Hilbert Spaces.
\newblock {\em Phys. Rev. Lett.}{ \bf 100}, 210503 (2008).

\bibitem{PhysRevA.68.050301}
Torres, J.~P., Alexandrescu, A., \& Torner, L.
\newblock Quantum spiral bandwidth of entangled two-photon states.
\newblock {\em Phys. Rev. A}{ \bf 68}, 050301 (2003).

\end{thebibliography}
\end{document}